\documentclass[
twocolumn,
superscriptaddress,
nofootinbib,
aps,
longbibliography,
pra,
floatfix,
]{revtex4-1}
\usepackage{xr-hyper}
\usepackage{amsmath}
\usepackage{amssymb}
\usepackage{graphicx}
\usepackage{braket}
\usepackage{color}
\usepackage{xcolor}
\usepackage{siunitx}
\usepackage{upgreek}
\usepackage[utf8]{inputenc}
\usepackage{physics}
\usepackage{multirow}
\usepackage{array}
\usepackage{url}
\usepackage{dcolumn}
\usepackage{bm}
\usepackage{hyperref}
\usepackage{natbib}
\usepackage{setspace}

\makeatletter
\newcommand*{\addFileDependency}[1]{
	\typeout{(#1)}
	\@addtofilelist{#1}
	\IfFileExists{#1}{}{\typeout{No file #1.}}
}
\makeatother

\newcommand*{\myexternaldocument}[1]{%
	\externaldocument{#1}%
	\addFileDependency{#1.tex}%
	\addFileDependency{#1.aux}%
}

\myexternaldocument{Supplementary}

\begin{document}



\title{Coherent spin dynamics of hyperfine-coupled vanadium impurities in silicon carbide}

\author{Joop Hendriks}
\thanks{These authors contributed equally to this work.}
\affiliation{Zernike Institute for Advanced Materials, University of Groningen, NL-9747AG  Groningen, The Netherlands}
\author{Carmem M. Gilardoni}
\thanks{These authors contributed equally to this work.}
\affiliation{Zernike Institute for Advanced Materials, University of Groningen, NL-9747AG  Groningen, The Netherlands}
\author{Chris Adambukulam}
\affiliation{School of Electrical Engineering and Telecommunications, UNSW Sydney, New South Wales 2052, Australia}
\author{Arne Laucht}
\affiliation{School of Electrical Engineering and Telecommunications, UNSW Sydney, New South Wales 2052, Australia}
\author{Caspar~H.~van~der~Wal}
\email{c.h.van.der.wal@rug.nl}
\affiliation{Zernike Institute for Advanced Materials, University of Groningen, NL-9747AG  Groningen, The Netherlands}

\date{Version of \today}


\begin{abstract}
Progress with quantum technology has for a large part been realized with the nitrogen-vacancy centre in diamond. Part of its properties, however, are nonideal and this drives research into other spin-active crystal defects. Several of these come with much stronger energy scales for spin-orbit and hyperfine coupling, but how this affects their spin coherence is little explored. Vanadium in silicon carbide is such a system, with technological interest for its optical emission at a telecom wavelength and compatibility with semiconductor industry. Here we show coherent spin dynamics of an ensemble of vanadium defects around a clock transition, studied while isolated from, or coupled to neighbouring nuclear spins. We find spin dephasing times up to 7.2~$\upmu$s, and via spin-echo studies coherence lifetimes that go well beyond tens of microseconds. We demonstrate operation points where strong coupling to neighbouring nuclear spins does not compromise the coherence of the central vanadium spin, which identifies how these can be applied as a coherent spin register. Our findings are relevant for understanding a wide class of defects with similar energy scales and crystal symmetries, that are currently explored in diamond, silicon carbide, and hexagonal boron nitride.
\end{abstract}

\maketitle

Spin-active color centres in solid state materials form a versatile class of qubits, offering a spin-photon interface in a scalable platform \cite{Majety2021c, Wan2020, Castelletto2020a}. Each combination of defect and host material, like diamond, SiC, hexagonal BN, and Si, comes with unique properties and advantages. The best-known example is the NV center in diamond, with long coherence times even at room temperature \cite{Herbschleb2019}. The NV center also enables the control of environmental nuclear spins, which form highly coherent hybrid registers that can be used to extend the useful lifetime of the qubits through error correction protocols or decoherence protected subspaces \cite{Bradley2019, Bartling2022}. Despite the success of the NV center in diamond, some properties such as the emission wavelength and the Debye-Waller (DW) factor, are nonideal for applications \cite{Awschalom2018}.

Alternative systems like the silicon-vacancy \cite{Lekavicius2022, Carter2015, Anderson2019, Nagy2019} and the divacancy \cite{Koehl2011, Christle2015, Anderson2022} in SiC also show long coherence times at room temperature, with the added benefits of optical emission closer to the telecom wavelengths and a higher DW factor. Additionally, SiC is easier to process into advanced electronic and photonic structures than diamond. Semiconductor doping is well developed and it is commercially available on the wafer-scale. This enables device integration of the defects \cite{Bosma2022, Lukin2020, Babin2022, Anderson2019, Widmann2019}. These results show that investigating other combinations of host materials and impurities is worthwhile.

A largely unexplored but interesting class of defects is formed by systems with heavy elements. This brings high spin-orbit coupling, and in many cases also strong hyperfine interaction \cite{Debroux2021, Kobayashi2021, Diler2020}. Due to the stronger interaction with a range of environmental factors, these systems are rich in physics, and give rise to hybrid multi-qubit platforms with varied readout and control possibilities \cite{Gilardoni2021, Asaad2020}.
At the same time, many relevant novel two-dimensional semiconductors and insulators host interesting optically-active spin defects but are intrinsically subject to high spin-orbit coupling or nuclear-spin-rich environments \cite{Liu2022hBN, Gottscholl2021, Tsai2022, Ye2019}.
For SiC, there are several examples of defects formed by heavy elements that show promising spin lifetimes and coherences at cryogenic temperatures, such as Mo \cite{Bosma2018, Gilardoni2019} and Cr \cite{Diler2020} impurities. The neutrally charged V defect also exhibits long spin lifetimes \cite{Astner2022}, along with a rich hyperfine structure and emission in the telecom O-band \cite{Wolfowicz2020, Spindlberger2019a}. These properties do not only make the V defect an attractive system for future applications, but also an interesting platform to investigate the interplay of different energy scales and how they influence defect spin-coherence properties.

In this work, we explore the coherent spin dynamics of an ensemble of V color centers in a commercial 4H-SiC sample with natural abundance of isotopes. We perform optically-detected magnetic resonance (ODMR) measurements, with the ground-state spin system tuned to a clock transition with an applied magnetic field. Depending on the optical wavelength $\lambda$, we probe subensembles with and without nearby environmental nuclear spins, revealing that the V centers couple strongly to neighboring $^{29}$Si nuclear spins. We determine the dephasing time ($T_2^*$) from Ramsey interference measurements, and find that it depends on the configuration of neighboring environmental nuclear spins, with the longest observed dephasing time being 7.2~$\upmu$s. These long ensemble-averaged $T_2^*$ times are particularly important since dynamical decoupling sequences have reduced effectiveness for systems with anisotropic coupling to magnetic fields \cite{Merkel2021}. Finally, we study the time evolution of the Hahn spin-echo, which does not show any sign of decay in our measurement range up to 26~$\upmu$s free precession time. This indicates that the decoherence time $T_2$ is at least an order of magnitude longer, and that the strong coupling to nuclear spins does not necessarily result in degraded coherence properties.

\begin{figure}[ht!]
	\centering
	\includegraphics[width=\linewidth]{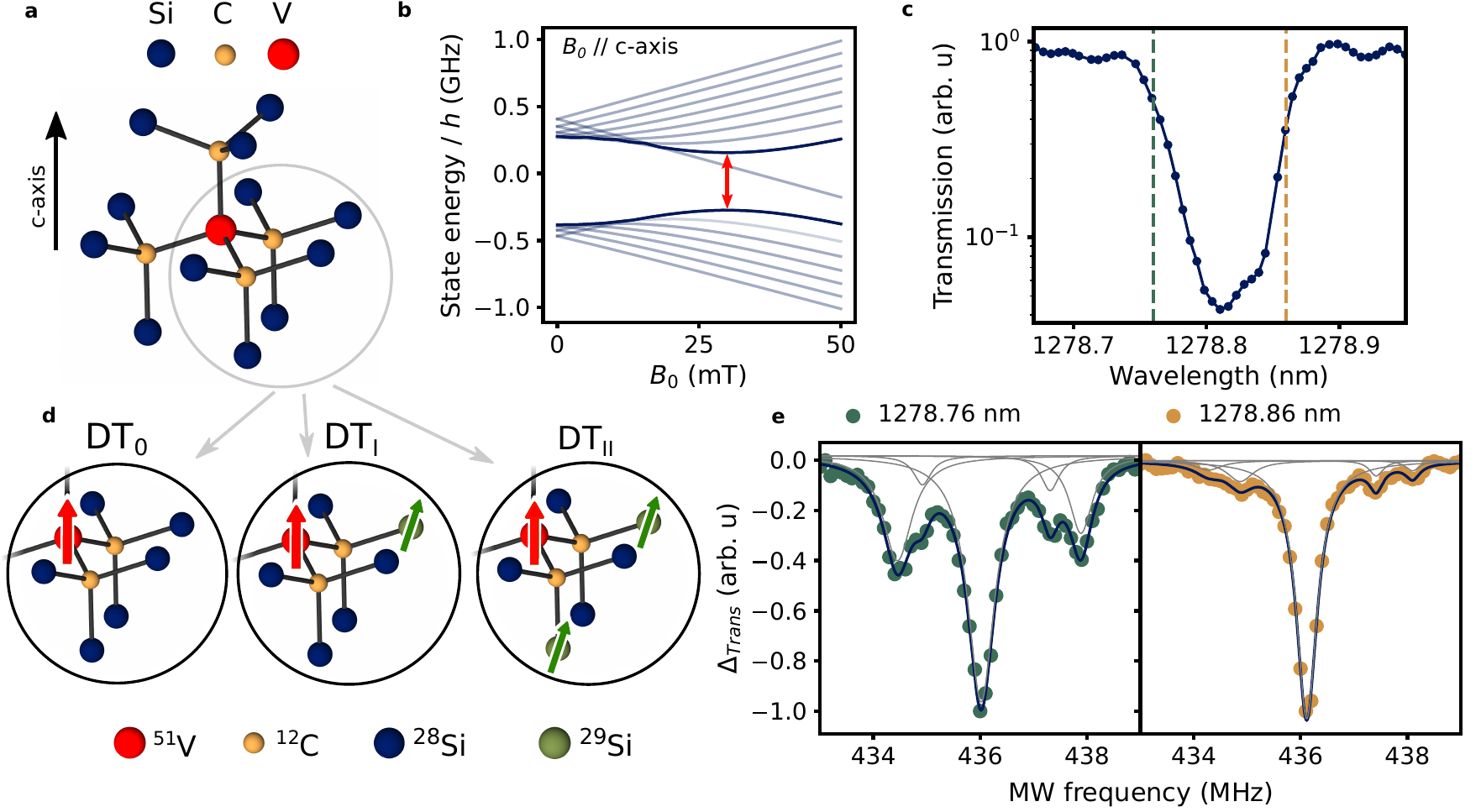}
	\caption{\textbf{Crystal structure, and optical and microwave spectroscopy of spin-active vanadium defects in 4H-SiC} \textbf{a}, Crystal structure of 4H-SiC with a V atom substituting a Si atom. \textbf{b}, Calculated state energies of the hyperfine sublevels of the V-defect ground state as a function of magnetic field $B_0$ applied parallel to the c-axis. The red arrow indicates the microwave (MW) transition that we focus on, which is a clock transition at $B_0 = 30~{\rm mT}$. \textbf{c}, Optical transmission spectrum of a V ensemble around resonant excitation between the ground state and a discrete optically excited state, with moderate inhomogeneity for this transition. The dashed lines indicate the optical wavelengths that are used for probing in panel \textbf{e}. \textbf{d}, Crystal structures as the lower-right part of panel \textbf{a}, illustrating how the V ensemble comprises of subensembles in which the individual V defects have zero, one, or a pair of nearest-neighboring $^{29}$Si (arrows represent nonzero nuclear spin). We denote these as DT$_0$, DT$_{{\rm I}}$ and DT$_{{\rm II}}$, respectively. \textbf{e}, ODMR spectra at $B_0 = 33~{\rm mT}$ obtained with 10~dBm MW power applied to the antenna. The spectrum obtained with probing wavelength of $\lambda$ = 1278.76~nm shows two sidepeaks on either side of a central peak. The central peak is due to the DT$_{{\rm 0}}$ subensemble. The sidepeaks are due to DT$_{{\rm I}}$ and DT$_{{\rm II}}$, for which the hyperfine coupling between the V spin and environmental nuclear spins causes a shift in transition energy.
When shifting the probing wavelength to $\lambda$ = 1278.86~nm, these sidepeaks are only weakly present since this causes less optical interaction with DT$_{{\rm I}}$ and DT$_{{\rm II}}$ relative to interaction with DT$_{{\rm 0}}$.
The dots are experimental results, the lines represent the summed (blue) and individual (grey) contributions to a fit with five Gaussian line shapes. All data is acquired at 2~K.}
	\label{fig:1}
\end{figure}

Figure~\ref{fig:1}a shows how a V impurity can substitute a Si atom in the SiC lattice, forming a V defect. In the 4H polytype this can take place at two inequivalent sites, where the difference in symmetry of the defect's environment causes a shift in emission wavelengths. In this study, we focus on the $\alpha$-line: the spectral line associated with the highest-energy zero-phonon line of these two sites \cite{Spindlberger2019a, Wolfowicz2020}. Due to the difference in valence electrons between Si and V, the substitutional V defect has a strongly localized single electron that originates from its 3$d$ shell. Interaction with the crystal field and spin-orbit coupling splits its energy levels into (in order of increasing energy) two ground-state Kramers' doublets (GS1 and GS2) and three excited-state doublets (ES1, ES2, ES3). The ground state GS1 has an effective electronic spin $S_{\rm{eff}} = 1/2$ with its quantization axis strongly pinned along the c-axis of the crystal \cite{Bosma2018,Gilardoni2019}. This effective spin has strong hyperfine interaction with the $I = 7/2$ nuclear spin of the $\sim$100$\%$ naturally abundant $^{51}$V isotope, giving rise to 16 hyperfine levels for GS1. The ground state GS1 can be described by a Hamiltonian of the type
\begin{equation}
	H_V = - \mu_B \mathbf{B}_0 \cdot \mathbf{g} \cdot \mathbf{S} + \mathbf{S} \cdot \mathbf{A} \cdot \mathbf{I}_V + \mu_N g_V \mathbf{B}_0 \cdot \mathbf{I}_V
	\label{eq:H_free}
\end{equation}
\noindent where $\mathbf{S}$ and $\mathbf{I}_V$, respectively, account for the electronic effective-spin and nuclear spin operators, $g_V$ and $\mathbf{g}$ are the vanadium-nucleus and electronic effective-spin g-factors, $\mathbf{B}_0$ is the static magnetic field and $\mathbf{A}$ is the hyperfine coupling tensor between the vanadium nucleus and the electron effective-spin. Using parameters from literature \cite{Tissot2021a, Wolfowicz2020, Astner2022}, slightly modified to describe our data (Supplementary Information), we obtain the energy levels illustrated in Fig.~\ref{fig:1}b. When increasing the magnetic field in the range up to $\sim$30~mT, the competition between the electronic Zeeman and hyperfine interaction results in several anti-crossings for the GS1 levels. The last one occurs at approximately 30~mT for the levels highlighted in dark blue in Fig.~\ref{fig:1}b. In our experiments, we focus on this particular clock transition, which involves energy levels that are linear combinations of states $\ket{\uparrow, -5/2}$  and $\ket{\downarrow, -7/2}$ (using the notation $\ket{m_{s_{eff}}, m_i}$, with spin projection along the c-axis). In these states, the electronic effective-spin and the V nuclear spin are always entangled, such that from now on we address these hyperfine-coupled states as the levels of the vanadium spin system (below we will address how, in turn, this system has hyperfine interaction with nuclear spins in its environment).

Figure~\ref{fig:1}c shows the optical transmission spectrum, measured around resonance with the GS1-ES1 transition (experimental techniques are detailed in Methods and the Supplementary Information). Our subsequent experiments are based on microwave (MW) magnetic resonance on GS1, which is probed optically by detecting the transmission of a laser beam with wavelength $\lambda$ (ODMR). In order to detect sufficient signal, we measure the transmission at two wavelengths (1278.76~nm and 1278.86~nm) away from the transmission minimum, as indicated by the dashed lines in Fig.~\ref{fig:1}c. Besides probing the population in the hyperfine levels, the near-resonant laser also plays an important role for creating a population imbalance between them.

The ODMR spectra measured at 33~mT depend strongly on the optical wavelength used for probing, as shown in Fig.~\ref{fig:1}e. For $\lambda$ = 1278.76~nm, the ODMR spectrum shows a central strong peak accompanied by two sidepeaks on either side. In contrast, for $\lambda$ = 1278.86 these sidepeaks are much weaker, and the spectrum mainly consists of the central magnetic resonance line. Sidepeaks in ODMR spectra are commonly found when electronic states couple to neighboring nuclear spins through hyperfine interaction \cite{Lekavicius2022, Carter2015}. In naturally isotopic SiC, the probability of finding a V defect with a single $^{13}$C in the nearest-neighbor shell is only $1.5 \%$. In contrast, defects with a single or a pair of $^{29}$Si isotopes in the nearest-neighbor shell for Si occur with probabilities $22 \%$ and $6 \%$, respectively (Fig.~\ref{fig:1}d). The additional mass of these isotopes with respect to the more abundant $^{28}$Si and $^{12}$C counterparts increases the optical transition energy of the defects that they are adjacent to via the isotope effect \cite{Wolfowicz2020}. Due to the large shift in resonance frequency and low abundance, we do not consider subensembles with a $^{13}$C isotope. We assign the sidepeaks present in the spectrum of Fig.~\ref{fig:1}e(left) to magnetic-resonance transitions between hybrid states formed by the central vanadium system coupled to a single environmental $^{29}$Si spin (inner sidepeaks) or to a pair or environmental $^{29}$Si spins (inner and outer sidepeaks). These results demonstrate that it is possible to almost exclusively address a subensemble which only couples to $^{28}$Si by using a long wavelength for optical probing ($\lambda$ = 1278.86~nm), and a subensemble with defects coupled to a single or a pair of $^{29}$Si by using a short wavelength ($\lambda$ = 1278.76~nm).

In order to more accurately characterize the coupling between the vanadium system and the neighboring nuclear spins, we measured the ODMR spectra as a function of the magnetic field around the clock transition for both optical detection wavelengths, as shown in the left panels of Fig.~\ref{fig:2}. The right panels show simulated data based on a model that uses the Hamiltonian of Eq.~\ref{eq:H_free} modified to include the quadrupole energy associated with the V nucleus \cite{Pooransingh2006,Nagasawa1964} and the interaction with the environmental $^{29}$Si spins (see Methods).
When probing with $\lambda$ = 1278.86~nm, we consistently find a narrow single peak, indicating that we almost exclusively detect defects with only $^{28}$Si nuclear spins in the second nearest neighboring shell. In contrast, for $\lambda$ = 1278.76~nm we observe sidepeaks at varying energies away from the central peak due to strong hyperfine coupling between the V electronic spin and the $^{29}$Si spins in the environment. Based on the model, we extract an absolute coupling $A = (A_\parallel^2 + A_\bot^2)^{1/2} = 9.7 \pm 0.5$~MHz. Using the relative height of the peaks in the spectra, we estimate the composition of each ensemble addressed optically in terms of subensembles with zero, one or two neighboring $^{29}$Si. From this, we find that for $\lambda$ = 1278.76~nm we predominantly address V defects with two neighboring $^{29}$Si spins, with subensembles with a single or no neighboring $^{29}$Si spin playing a smaller role, with a relative ratio of 3:1:1 (see Supplementary Information).

For both optical probing wavelengths at $B_0 < 28~{\rm mT}$ we see an additional MW-resonance peak, for which the MW resonance frequency depends linearly on the magnitude of the static magnetic field (see also Fig.~\ref{fig:1}b). The MW transition associated with this peak, between the levels $\ket{\uparrow, -7/2}$ and $\ket{\uparrow, -3/2}$, involves a $\Delta m_i = 2$ nuclear spin flip for the vanadium nuclear spin. Thus, in our measurement geometry, antenna-induced electric or magnetic fields can drive pure vanadium nuclear spin flips \cite{Gilardoni2021}, although we were not able to observe Rabi oscillations between these levels.

\begin{figure}[ht!]
	\centering
	\includegraphics[width=\linewidth]{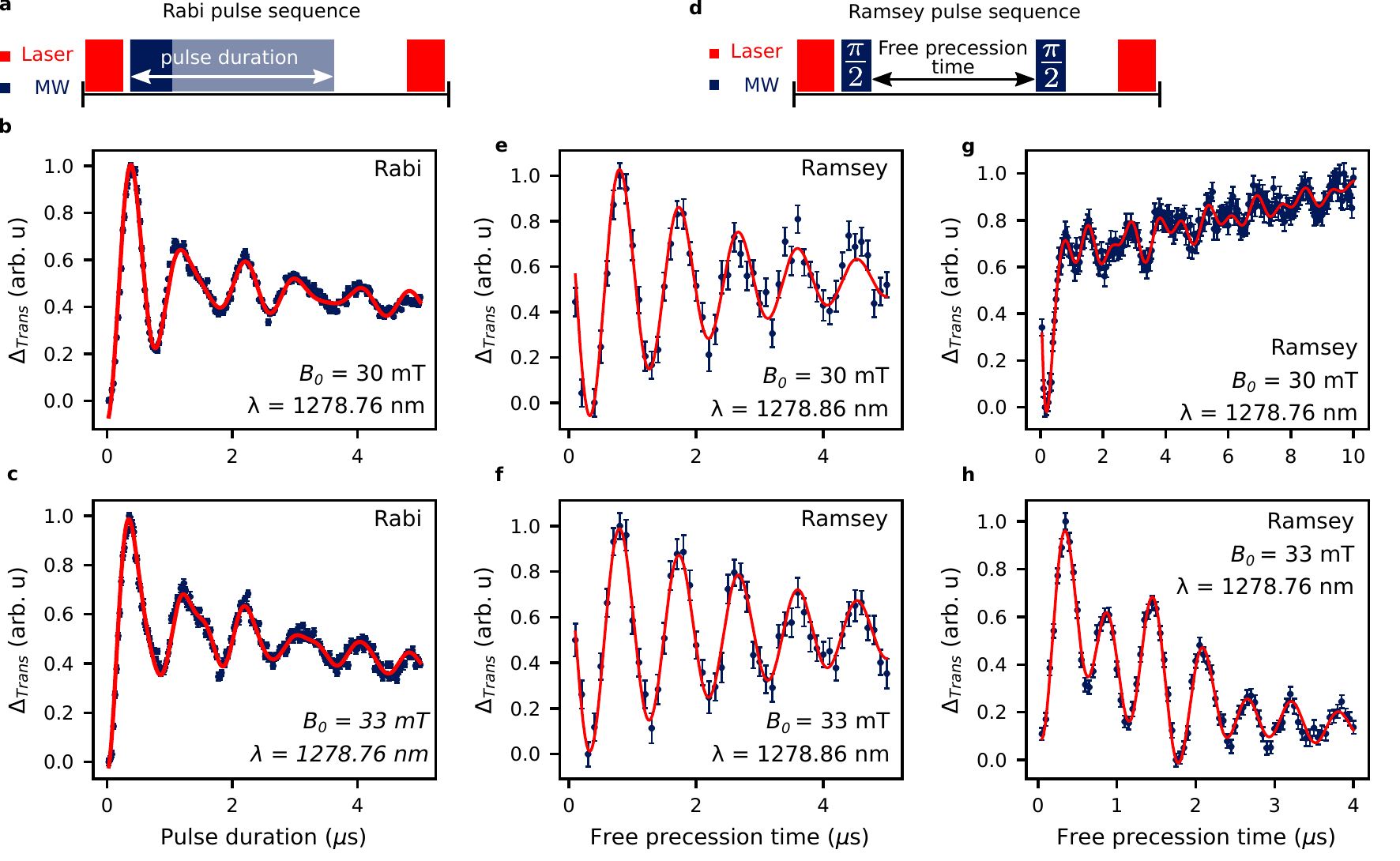}
	\caption{\textbf{Variations in sample transmission as a function of pulsed microwave driving, showing Rabi oscillations and Ramsey interference.} \textbf{a}, Pulse sequence for measuring Rabi oscillations. \textbf{b, c}, Rabi oscillations at $B_0 = 30~{\rm mT}$ (\textbf{b}) and $B_0 = 33~{\rm mT}$ (\textbf{c}) with optical probing at $\lambda = 1278.76$~nm. \textbf{d}, Pulse sequence for measuring Ramsey interference fringes. \textbf{e-h} Ramsey fringes with $\lambda$ = 1278.86~nm  (\textbf{e, f}) and $\lambda$ = 1278.76~nm (\textbf{g,h}) for $B_0 = 30$ and 33~mT. The applied power to the antenna is 25~dBm, while the MW frequency is detuned from the central ODMR peak by 1~MHz in \textbf{e - g} and on resonance in \textbf{h}. The error bars represent the 95\% confidence interval of the mean. All data is acquired at 2~K.}
	\label{fig:3}
\end{figure}

We characterize the coherence properties of our ensembles with pulsed ODMR, giving time-resolved measurements. Figure~\ref{fig:3}a-c show the Rabi pulse-sequence and Rabi oscillations obtained by probing the system with a wavelength of $\lambda$ = 1278.76~nm. We apply MW pulses of varying length, where the MW frequency is resonant with the central ODMR peak. Given that the ODMR spectrum of Fig.~\ref{fig:1}e(left) is dominated by five different peaks, we should account for having five allowed magnetic-resonance transitions for the hybrid systems formed by the V defects and the neighboring $^{29}$Si spins, each with their particular detuning from the MW driving field. Phenomenological fitting confirms that good fits require a linear combination of five different terms for Rabi oscillations, with contributions at five different Rabi frequencies.
Within error we must assume identical dephasing times for all five transitions.
These fits give ensemble-averaged driven dephasing times of $1.8 \pm 0.2~\upmu$s (Fig.~\ref{fig:3}b, error is the 95\% confidence interval) and $2.1 \pm 0.2~\upmu$s (Fig.~\ref{fig:3}c).
Similar data for other experimental parameters are presented in the Supplementary Information.

We characterize the free-precession ensemble-averaged dephasing time $T^*_2$ from measurements with Ramsey pulse sequences (Fig.~\ref{fig:3}d). The MW-pulse lengths for $\pi/2$ flips were calibrated with the Rabi measurements.
Figures~\ref{fig:3}e-h show the results from the Ramsey interferometry at the clock-transition (30~mT) and slightly away from it (33~mT), for both optical wavelengths. We fit the Ramsey fringes to the phenomenological equation
\begin{equation}
	 \sum_{i=1}^{n}  e^{- \frac{t}{T_{2i}^*}} \cdot  a_i sin(2\pi f_i t + \phi_i),
     \label{eq:Ramseyfit}
\end{equation}
\noindent where $n = 5$ when $\lambda$ = 1278.76 nm and $n = 1$ when $\lambda$ = 1278.86 nm. Here $a_i$ is the relative amplitude of the $i_{th}$ frequency component, $f_i$ is the detuning of the MW frequency with respect to transition $i$ (consistent with the ODMR results), and $\phi_i$ is a phase. For $n=5$, each contribution can have its own dephasing time $T_{2i}^*$, but as for the Rabi case we find for Figs.~\ref{fig:3}h that we must assume within error the same value for each contribution. This yields $T_2^* = 2.3 \pm 0.5~\upmu$s (Fig.~\ref{fig:3}e), $T_2^* = 3.2 \pm 0.6~\upmu$s (Fig.~\ref{fig:3}f), and $T_2^* = 2.0 \pm 0.3~\upmu$s (Fig.~\ref{fig:3}h).

Notably, fitting the results in Fig.~\ref{fig:3}g \textit{does} require that we use different $T_{2i}^*$ in Eq.~\ref{eq:Ramseyfit}. This signal clearly has contributions from five different precession frequencies, with frequency-specific dephasing time scales. These five different frequencies result from modulations of the V-system precession frequency due to the coherent dynamics of the neighboring $^{29}$Si nuclei. This phenomenon is denoted by electron-spin echo envelope modulation (ESEEM) \cite{Rowan1965, Seo2016}. We could only fit this data by considering at least two different exponential-decay timescales, that differ by almost an order of magnitude: the frequencies that are related to the central peak and the outer sidepeaks dephase in a characteristic time $T_{2}^* = 0.7 \pm 0.2~\upmu$s, while the frequencies that correspond to the inner sidepeaks dephase with a characteristic $T_{2}^* = 7.2 \pm 4.1~\upmu$s.

Different $T_2^*$ timescales depending on the presence of neighboring environmental spins and their initial state have been observed in other systems \cite{Onizhuk2021, Liu2022hBN}, both theoretically and experimentally. Although we cannot pinpoint the origin of these different timescales from our ensemble measurements, we can conclude that coupling to neighboring $^{29}$Si nuclear spins may extend the dephasing time of the V spin system. Single-defect experiments are required to confirm whether this behavior is related to a particular subensemble, or to the presence of nuclear spins in a particular initial state. For the measurement presented in Fig.~\ref{fig:3}g, subensembles with zero, one, or two $^{29}$Si neighboring the V defect contribute to the total signal. While only the subensemble with two neighboring $^{29}$Si contributes to the frequencies related to the outer sidepeaks, multiple subensembles may contribute to the frequencies related to the inner and central peaks. In our fits, by associating the dephasing time $T_{2}^*$ to particular frequencies, the contributions from different subensembles to the same frequency cannot be distinguished. A different approach to fitting the data based on ESEEM (see Supplementary Information) confirms the two timescales found above, but does not provide a conclusive physical explanation for the origin of the different dephasing timescales.

When probing with $\lambda$ = 1278.86~nm (Fig.~\ref{fig:3}e,f), where the ODMR results indicate that we address predominantly a subensemble with zero $^{29}$Si in the closest shell, we do not see a significant improvement of the dephasing time at the clock-transition point (30~mT). In fact, we measure the longest dephasing time at 33~mT, slightly away from the clock-transition point. At the clock-transition point, the eigenenergies of the levels involved are, to first order, insensitive to magnetic-field noise. However, for the V defect, the hyperfine coupling between the electronic effective-spin and the central V nuclear spin has a large contribution from the dipolar term and can thus be strongly affected by electric fields or strain \cite{Gilardoni2021,Tissot2021}. In this way, the inhomogeneous distribution of electrostatic environments throughout the ensemble could contribute significantly to dephasing \cite{Onizhuk2021}.

At 30~mT and with optical probing at $\lambda = 1278.76~$nm, the different frequencies contributing to the signal occur in steps of approximately 0.27~MHz. This reflects ESEEM modulations at frequencies compatible with the Larmor precession rate of a bare $^{29}$Si nuclear spin ($\sim$0.25~MHz at 30~mT). This indicates that, at the clock-transition point, the neighboring $^{29}$Si nuclear spins do not feel a large additional semi-classical field due to the presence of the V electronic spin \cite{Childress2006}. This observation also explains the suppression of the sidepeaks observed in the spectra of Fig.~\ref{fig:2}b between 29 and 30~mT. A similar effect is observed for spin-1 systems like the NV center in diamond or the divacancy in SiC, 
where the hyperfine coupling to environmental spins can be approximately turned off by bringing the electronic system into an eigenstate with zero electron-spin projection ($m_s=0$ state). In these systems, this is an important feature enabling electronic-spin mediated gates on environmental nuclear spins \cite{vanderSar2012, Childress2006}.

\begin{figure}[ht!]
	\centering
	\includegraphics[width=0.5\linewidth]{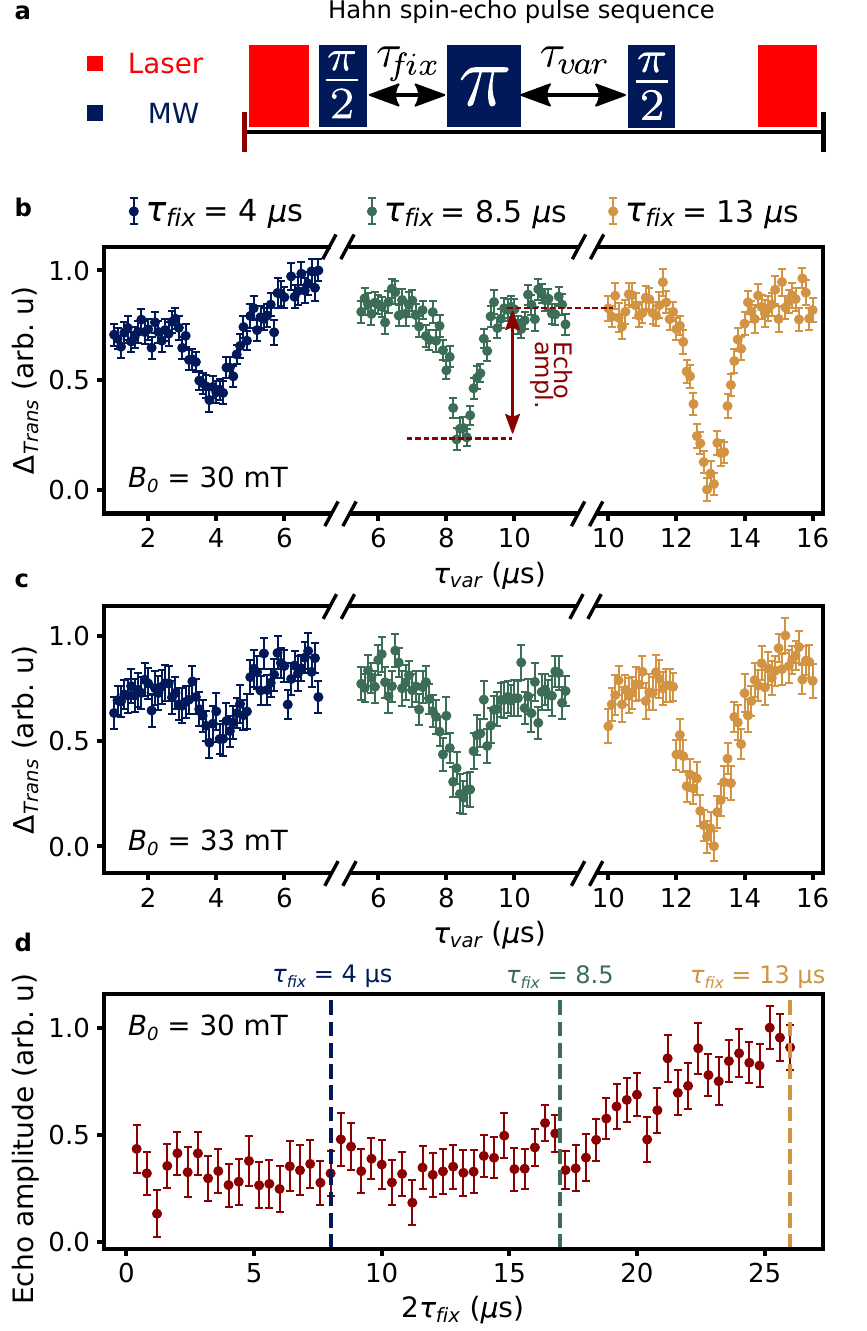}
	\caption{\textbf{Hahn spin-echo measurements as a function of the free spin precession time.} \textbf{a}, Pulse sequence for generating Hahn spin echoes, where the total free precession time is $\tau_{fix} + \tau_{var}$ . \textbf{b, c}, Hahn spin echoes measured at 30~mT and 33~mT for several values of $\tau_{fix}$, probed with $\lambda$ = 1278.76~nm light. \textbf{d}, Amplitude of the Hahn echo as a function of the total free precession time 2$\tau_{fix}$, with $\tau_{var}=\tau_{fix}$ in the pulse sequence (directly measured, other run than data in \textbf{b}). The amplitude is defined as the difference in signal measured at $\tau_{var}=\tau_{fix}$ and $\tau_{var}=\tau_{fix}$~+~3~$\upmu$s.
	The error bars represent the 95\% confidence interval of the mean. All data is acquired at 2~K with 25~dBm applied MW power.} 	
	\label{fig:4}
\end{figure}

Finally, we apply a Hahn spin-echo pulse sequence (Fig.~\ref{fig:4}a) to probe the decoherence time $T_2$ of the V ensemble, where the MW frequency is tuned to the central ODMR peak. Figure~\ref{fig:4}b,c show the Hahn spin echoes at $B_0 = 30$ and $33$~mT and $\lambda = 1278.76$~nm, corrected for setup-induced fluctuations (details in Supplementary Information). For both magnetic fields, we observe similar behavior in shape and amplitude of the echoes. The echo dip \textit{increases} in amplitude and becomes more symmetric as we increase $\tau_{fix}$ from 4 to 13~$\upmu$s. Figure~\ref{fig:4}d shows the echo amplitude, defined as the difference between the signal measured at $\tau_{var} = \tau_{fix}+3~\upmu$s and at $\tau_{var} = \tau_{fix}$.
Limited by the sensitivity of our optical readout, we could only measure up to free precession time $2\tau_{fix} = 26~\upmu$s. Notably, up to this timescale, rather than showing a decay, the echo amplitude \textit{increases}. This observed increase is likely a result of ESEEM modulation due to the presence of neighboring nuclear spins \cite{Rowan1965, Seo2016}. The specific shape of the ESEEM envelope depends on the particular anisotropy of the hyperfine coupling tensor \cite{Childress2006}, preventing a quantitative analysis of the ESEEM lineshape from our ensemble-averaged data. We can, however, qualitatively reproduce the measured lineshapes (see Supplementary Information). We thus determine that $T_2$ is at least an order of magnitude longer than 26~$\upmu$s.

In conclusion, these results show strong 9.7~MHz coupling of the vanadium spin system to neighboring $^{29}$Si nuclear spins, along with promising coherence properties. By probing at different regions of the optical spectrum, the relative contribution of subensembles with different numbers of neighboring nuclear spins to the ODMR spectra can be changed. This is also reflected by the Ramsey interference results, where only a single frequency can describe the observed data when we probe with optical pulses with $\lambda$ = 1278.86~nm, while we observe multiple frequencies when using $\lambda$ = 1278.76~nm. Moreover, there are at least two dephasing timescales required to phenomenologically describe the data, with the long timescale $T_2^*=7.2~\upmu$s being a factor 10 longer than the short timescale. By applying a Hahn-echo pulse sequence, we observed that the coherence time $T_2$ is significantly longer than 26~$\upmu$s (our detection scheme does not allow for probing it at longer timescales).

Our results demonstrate that complex electronic structures with several degrees of freedom are not necessarily detrimental to defect stability and coherence. With relevance for a broad class of defects in Si \cite{Higginbottom2022}, diamond \cite{Debroux2021}, hBN \cite{Haykal2022, Liu2022hBN, Gottscholl2021}, SiC and other material platforms, our results on V defects show that substantial electron-spin coherence is possible provided that orbital degrees of freedom are frozen. For the V system specifically, our results demonstrate a series of properties that place it as a strong contender on a par with the NV center in diamond. We have shown that the coupled electron-nuclear spin system is long-lived and can be detected optically even at the ensemble level. Our estimates indicate that the single-defect spin-based optical contrast could reach several tens of percents (Supplementary Information). Additionally, the presence of hyperfine levels with specific couplings to neighboring $^{29}$Si enables electron-spin mediated gates of environmental nuclear spins, with a clear path towards establishing high-density quantum-state storage clusters with a telecom-compatible optical interface. The observation that we can drive $\Delta m_i = 2$ transitions using microwaves indicates an important role for direct or indirect electric-field-driving of the V nuclear spin, with a clear path towards electric-field control of nuclear spin coherences \cite{Asaad2020,Gilardoni2021,Savytskyy2022}. Finally, we propose that inhomogeneous electric fields could be a significant source of dephasing for our ensemble. This effect can however be significantly reduced by moving into the single-emitter regime, where device-enabled control of the electrostatic environment is also possible \cite{Widmann2019, Anderson2019}. 

\section{Methods}
In all experiments we used a commercial 4H-SiC sample with a V defect concentration of $10^{17}~cm^{-3}$, cooled inside a cryostat to 2~K. The geometry of the setup is such that the static $B_0$ field, the oscillating $B_1$ field and the $k$-vector of the laser beam are parallel to the crystal c-axis of the sample. Depending on the measurement, we apply different optical pulses to initialize and read out the defects, and MW pulse sequences to drive the transition between the two ground states (see Supplementary Information for specific pulse sequence per measurement). We collect the light that is transmitted through the sample with a tunable-gain photodiode, which in turn is connected to a lock-in amplifier. In this way, the obtained signal $\Delta_{trans}$ is the difference in transmission between driving the defects with microwaves, and not driving them (except for the resonant excitation spectroscopy data, which is the difference between optically exciting the sample and leaving it in the dark). The $B_1$ field is generated with a loop antenna which is attached to the sample, where the $R = $50~$\upmu$m radius hole allows for the laser beam to go trough. The antenna creates a homogeneous field in the plane of the loop, but decreases with $\frac{R^2}{(z^2 + R^2)^{3/2}}$ with $z$ defined along the c-axis of the sample. As a consequence, only approximately 10\% of the defects that are addressed optically are driven by the $B_1$ field in a $500~\upmu m$ thick sample, thereby decreasing the contrast that could be achieved by a thinner sample (see Supplementary Information for details about contrast). This becomes a limiting factor when the optical pulses are far apart, such as in the Hahn-echo measurements, allowing for a maximum separation between the optical pulses of 30~$\upmu$s.

\begin{table}[ht!]
	\caption{Coupling parameters used for simulations in Fig.~\ref{fig:2}	 \label{tab:Parameters}}
	\begin{tabular}{|c|c|}
		\hline
		\multicolumn{2}{|c|}{V-system parameters}\\ \hline
		$g_\parallel, g_\bot$ & 1.664,0\\
		$A_{V,\parallel}, A_{V,\bot}$~(MHz) & -232.02,-162.32\\
		$Q_{zz}$~(MHz) & -0.2 \\ \hline
		\multicolumn{2}{|c|}{V-Si coupling parameters}\\ \hline
		$A_{Si,\parallel}, A_{Si,\bot}$~(MHz) & -8.2, -3.6\\ \hline
	\end{tabular}	
\end{table}

\begin{figure}[ht!]
	\centering
	\includegraphics[width=0.5\linewidth]{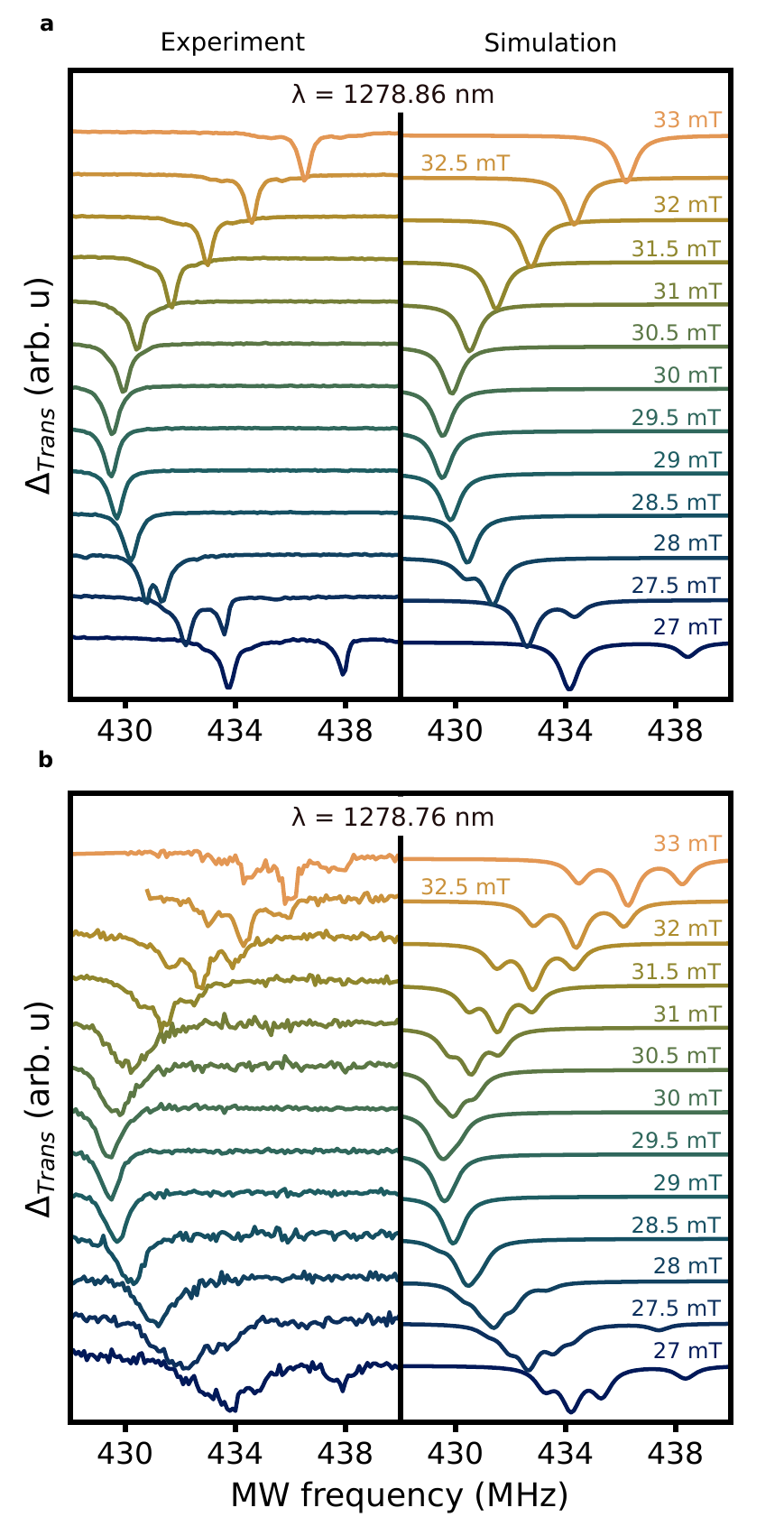}
	\caption{\textbf{Experimental (left panels) and simulated (right panels) ODMR spectra as a function of microwave (MW) frequency, for values of applied magnetic field $\boldsymbol{B_0}$ as labeled.}  \textbf{a}, ODMR spectra probed with 1278.86~nm light. The simulated traces are for an ensemble without neighboring nuclear spins. \textbf{b}, ODMR spectra probed with 1278.76~nm light. The simulated traces are for an ensemble with two $^{29}$Si nuclear spins in the nearest-neighbor shell of Si atoms. All individual traces are normalized (to strongest dip in the transmission signal) and offset by 0.8 for clarity. Data acquired at 2~K and 10~dBm MW power applied to the antenna.}
	\label{fig:2}
\end{figure}

For simulating the ODMR spectra presented in Fig.~\ref{fig:2}, we modified the Hamiltonian in Eq.~\ref{eq:H_free} to include the quadrupole energy of the V nucleus, as well as the energy contributions from the neighboring $^{29}$Si nuclear spins. This gives rise to a Hamiltonian of the kind
\begin{equation}
\begin{split}
H = &H_V + Q_{zz} I_{V,z}^2 + \\ &\sum_{k=1}^{2} S_z (A_{k,\parallel}  I_{k,z} + A_{k,\bot} I_{k,\bot}) + \upmu_N g_{Si} \mathbf{B}_0 \cdot \mathbf{I}_k
\end{split}
\label{eq:H}
\end{equation}
\noindent where $Q_{zz}$ is the strength of the quadrupole interaction experienced by the V nucleus due to an electric-field gradient in the $z$ direction \cite{Asaad2020}. Due to the axial symmetry of the defect, this term is expected to dominate the quadrupole energy contribution, and must be taken into account in order to simultaneously reproduce the magnetic-field dependence of all transitions observed in the spectra of Fig.~\ref{fig:2}. The summation in the third term introduces the energy contributions from up to 2 $^{29}$Si in the second-nearest neighbor shell, each with nuclear spin $I_k$. The nuclear Zeeman term is governed by the $^{29}$Si g-factor, $g_{Si}$. Additionally, we include the interaction between the electronic spin and each $k$-th $^{29}$Si nucleus due to hyperfine coupling with hyperfine parameters $A_{k,\parallel}$, $A_{k,\bot}$. We take in this case the semi-secular approximation, where we only consider terms that couple the $^{29}$Si nuclear spins to $S_z$. This is valid since the electron effective-spin is strongly pinned along the $z$ direction (i.e. c-axis) due to symmetry and spin-orbit coupling \cite{Gilardoni2019, Csore2020}. Finally, we note that in our experiments we actually address millions of defects and sample from every different possible arrangement of V and neighboring $^{29}$Si. Here, in order to minimize the number of fitting parameters, we approximate the contributions of all defects with the same number of neighboring $^{29}$Si with a single Hamiltonian. By comparing the microwave resonance frequencies obtained in experiments to those predicted based on this model (see Supplementary Information), we obtain the fitting parameters presented in Tab.~\ref{tab:Parameters}

\vspace{1cm}
\noindent \textbf{Acknowledgements}\\ We thank A.~Gali, and B.~Tissot for discussions, and M.~Trupke for discussions and sharing of SiC material. We are also grateful for technical support by H.~de~Vries, J.~Holstein, F.~van~der~Velde, and H.~Adema. Financial support was provided by the Zernike Institute BIS program, and the EU H2020 project QuanTELCO (862721). C.A acknowledges support from the University of New South Wales Scientia Program.

\vspace{1cm}
\noindent \textbf{Author Contributions}\\
The project was initiated by C.H.W. and C.M.G. Experiments were performed by C.M.G. and J.H. Data analysis and manuscript writing was performed by J.H., C.M.G, and C.H.W. C.A. and A.L. designed and provided the microwave antenna. J.H. and C.M.G. are co-first author. All authors read and commented on the manuscript.

\newpage

\onecolumngrid

\newpage

\newpage


\newpage

\twocolumngrid

\newpage


%

\end{document}